\documentclass[10pt,letterpaper,twocolumn]{article} 
\usepackage{ol2}
\usepackage[draft]{hyperref}
\usepackage{amsmath}
\usepackage[section]{placeins}
\usepackage{amsmath}
\usepackage{amssymb}
\usepackage{amsfonts}
\usepackage{dsfont}
\newcommand{\op}[1]{\widehat{#1}}            
\newcommand{\braket}[2]{\langle#1|#2\rangle} 
\newcommand{\ket}[1]{|#1\rangle}             
\newcommand{\bra}[1]{\langle #1|}            
\begin{document}

\twocolumn[ 

\title{Radial coherent and intelligent states of paraxial wave equation}

\author{Ebrahim Karimi,$^{1,*}$ and Enrico Santamato$^{1,2}$}

\address{
$^1$Dipartimento di Scienze Fisiche, Universit\`{a} di Napoli ``Federico II'', Complesso di Monte S. Angelo, 80126 Napoli, Italy
\\
$^2$Consorzio Nazionale Interuniversitario per le Scienze
Fisiche della Materia, Napoli  \\
$^*$Corresponding author: karimi@na.infn.it
}

\begin{abstract}
Ladder operators for the radial index of the paraxial optical modes in the cylindrical coordinates are calculated. The operators obey the su(1,1) algebra commutation relations. Based on this Lie algebra, we found that coherent modes constructed as eigenstates of the destruction operator or resulting from the action of the displacement operator on the fundamental mode are different. Some properties of these two kinds of radial coherent modes are studied in detail.
\end{abstract}


 ] 

\noindent Beside Linear Momentum (LM) and Spin Angular Momentum (SAM), photons possess transverse degrees of freedom: Orbital Angular Momentum (OAM) and radial profile distribution~\cite{arnold:08}. The light OAM received a lot of attentions in both classical and quantum regimes during last two decades~\cite{gibson:04,mair:01,nagali:09b,karimi:10a}. The light OAM is characterized by a vortex optical phase, whose topological charge is given by the azimuthal index $m=0,\pm1,\pm2,\dots$. On the contrary, the Radial Index (RI) $p=0,1,2,\dots$, which is associated to the intensity distribution of the light in the transverse plane, has been out of interest almost completely. Nevertheless, these two transverse degrees of freedom, i.e. OAM and radial intensity distribution, are strictly correlated, and different OAM generators produce specific (and different) distributions of radial modes. For instance, cylindrical mode converters generate pure Laguerre-Gauss (LG) modes starting from the Hermite-Gauss (HG) modes, spiral phase plates and pitch-fork holograms generate Hypergeometric-Gaussian (HyGG) modes, and $q$-plates generate other kind of radial HyGG modes as well~\cite{padgett:04,berry:04,bekshaev:08,karimi:07,karimi:09a}. Furthermore, some type of OAM projectors such as the single mode optical fibers widely used in the light quantum regime, affect the radial profile independently of the OAM value. In fact, optical fibers are not good OAM detectors since they are sensible only to specific radial indices~\cite{karimi:09b,molinaterriza:07a}. \\
Unlike LM and OAM photon eigenstates, the radial modes exhibits a more complicate structure, which render their manipulation less obvious. In this respect, to overcome the difficulties inherent to the radial profile of paraxial beans, a deeper understanding of the underlying symmetry is advisable. Such hidden symmetry, in fact, can be used to unveil novel and interesting features of the radial modes. In particular, the radial profile operator and \textit{its conjugate-variable} can be found from the symmetry of the state. Among the possible radial states, of particular interest is the coherent state(CS). The CS, in fact, is the quantum state closest to the classical one and provides a minimum uncertainty between two conjugate observables~\cite{perelomov:86}. Erwin Schr\"odinger was the first to derive a CS for the case of the harmonic oscillator in 1926, when he was looking for a quantum wave-packet with minimum momentum-position uncertainty~\cite{schrodinger:26}.
Later on, generalized coherent states (GCS) were developed for different dynamical systems and symmetries. In order to define the GCS, several strategies have been used: minimizing the uncertainty relations of two conjugate observables, thus creating \textit{ideal} states, finding the group lowering operator eigenvalue, thus creating \textit{Intelligent} States (IS), or introducing a suitable displacement operator acting on some reference state (ground state, usually), thus creating proper \textit{Coherent} States (CS)~\cite{perelomov:86,barut:71}.\\
In this paper, we studied the dynamical (or hidden) symmetry of the scalar paraxial wave equation, by considering the group algebra of the LG modes. Both IS and CS radial modes were then constructed starting from the algebra and some features of these modes were described in the position representation of the state.
The most common optical beams generated by laser sources obeys the scalar Paraxial Wave Equation (PWE)
\begin{eqnarray}\label{eq:pwe}
	\left(\mathbf{\nabla}^2_{\bot}+4i\partial_\zeta\right)\psi(\rho,\phi,\zeta)=0,
\end{eqnarray}
where $(\rho,\phi,\zeta)$ and $\mathbf{\nabla}^2_{\bot}$ are dimensionless cylindrical coordinate and the transverse Laplacian, respectively~\cite{note}.
The field at any given $\zeta$-plane can be calculated by applying a unitary propagator $\op{U}$ to the pupil wave-function, $\ket{\psi}_\zeta=\op{U}(\zeta,\zeta')\ket{\psi}_{\zeta'}$, where $\op{U}(\zeta,\zeta')=\exp{\left(\frac{i}{2}\,(\zeta-\zeta')\mathbf{\nabla}^2_{\bot}\right)}$.
In the position representation the operator $\op{U}$ is the Fresnel propagation kernel $K(\rho,\phi,\zeta;\rho',\phi',\zeta')=\bra{\rho,\phi}\op{U}(\zeta,\zeta')\ket{\rho',\phi'}$. Hereafter, without loss of generality, we consider the pupil solution of the PWE, and keep whole calculations in $\zeta=0$ plane.
LG modes, $\mbox{LG}_{p,m}(\rho,\phi,0):=\braket{\rho,\phi}{\mbox{p,m}}$, are a solutions of the scalar PWE with pupil function given by
\begin{figure}[t]
\begin{center}
	\includegraphics[scale=0.33]{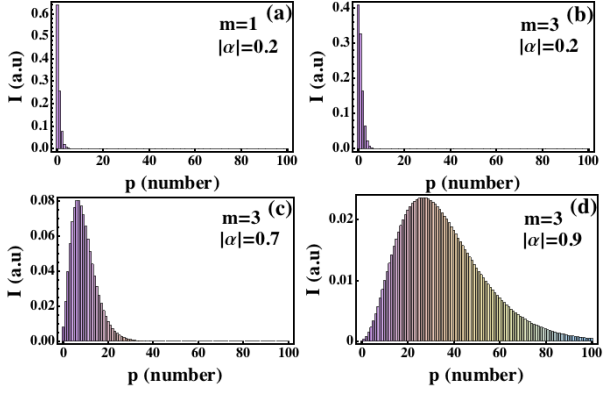}
	\caption{\label{fig:lg_basis} (Color online) Simulated spectra of different RCS in the basis of LG modes. (a) $m=1$, $|\alpha|=0.2$ (b) $m=3$, $|\alpha|=0.2$ (c) $m=3$, $|\alpha|=0.5$, and (d) $m=3$, $|\alpha|=0.9$. 
}
\end{center}
\end{figure}
\begin{eqnarray}\label{eq:paraxial_lg}
	\mbox{LG}_{p,m}(\rho,\phi,0)=\sqrt{\frac{ 2^{|m|+1}p!}{\pi(p+|m|)!}}\,\rho^{|m|} e^{-\rho^2} L_{p}^{|m|}\left(2\rho^2\right),\nonumber
\end{eqnarray}
where $m$ is an integer number defining the OAM eigenvalue, $p\geq0$ is a not negative integer number defining the radial nodes in the beam transverse plane, and $L_{p}^{|m|}(.)$ is the generalized Laguerre polynomial. The LG modes are orthogonal set of PWE's solution, i.e. $\braket{\mbox{p}',\mbox{m}'}{\mbox{p,m}}=\delta_{p,p'}\delta_{m,m'}$, and carry a finite power~\cite{siegman:86}. Here, we focus our attention to the radial node Hilbert space keeping the OAM number $m$ fixed. In order to find the dynamical (or hidden) symmetry lying behind the radial nodes of LG modes, we need the algebra of the node space. A straightforward calculation shows that the radial nodes ladder operators for the LG modes are given by
\begin{eqnarray}\label{eq:ladder_def}
	\op{\cal{P}}_+\ket{\mbox{p,m}}&=&{\cal{P}}_+\ket{\mbox{p+1,m}}\cr
	\op{\cal{P}}_-\ket{\mbox{p,m}}&=&{\cal{P}}_-\ket{\mbox{p-1,m}}.
\end{eqnarray}
where $\op{\cal{P}}_{\pm}=\op{\cal{Q}}\mp\left(\frac{|m|}{2}-\rho^2\right)$, $\op{\cal{Q}}=\frac{1}{2}\rho\,\partial_\rho+2\rho^2-(\op{p}+|m|+1)$ are the raising and lowering operators with eigenvalues ${\cal P}_{+}=\sqrt{(p+1)(p+|m|+1)}$ and ${\cal P}_{-}=\sqrt{p(p+|m|)}$, respectively. It can be easily checked that the ladder operators obey the su(1,1) Lie algebra
\begin{eqnarray}\label{eq:algebra}
	[\op{\cal{P}}_+,\op{\cal{P}}_-]=-2\op{\cal{P}}_0\quad [\op{\cal{P}}_0,\op{\cal{P}}_\pm]=\pm\op{\cal{P}}_\pm,
\end{eqnarray}
where $\op{\cal{P}}_0=\left(2\op{p}+|m|+1\right)/2$. In the equations above, $\op{p}$ is the second-order differential operator
\begin{eqnarray}\label{eq:algebra}
	 \op{p}=-\frac{1}{8\rho}\partial_\rho(\rho\,\partial_\rho)+\left(\frac{|m|^2}{8\rho^2}+\frac{\rho^2}{2}-\frac{|m|+1}{2}\right).
\end{eqnarray}
The states $\ket{\mbox{p,m}}$ are eigenstates of $\op{p}$ with eigenvalue $p$, i.e. we have $\op{p}\,\ket{\mbox{p,m}}=p\ket{\mbox{p,m}}$. The operators acting in the $\zeta$-plane can be obtained by the corresponding operators in the pupil plane $\zeta=0$ from $\op{\cal{N}}_\zeta=\op{U}(\zeta,0)^{\dag}\op{\cal{N}}_0\op{U}(\zeta,0)$, where $\op{\cal{N}}$ is any operator in the node Hilbert space in the pupil plane. The Casimir operator is given by $\op{\cal{C}}=\op{\cal{P}}_0^2-1/2\left(\op{\cal{P}}_+\op{\cal{P}}_-+\op{\cal{P}}_-\op{\cal{P}}_+\right)$, where $\op{\cal{C}}\ket{\mbox{p,m}}=(m^2-1)/4\ket{\mbox{p,m}}$~\cite{perelomov:86}.
By knowing the underlying symmetry, one can try to exploit further properties of the radial node space. We here introduce the CS of the Lie group as the state resulting formally from the displacement of the ground state~\cite{perelomov:86}. Later, we will introduce also the IS as the eigenvector of the lowering operator~\cite{barut:71}. Unlike in the case of the harmonic oscillator, in the case of su(1,1) dynamics these two states are not coincident.\\
\begin{figure}[h]
\begin{center}
	\includegraphics[scale=0.33]{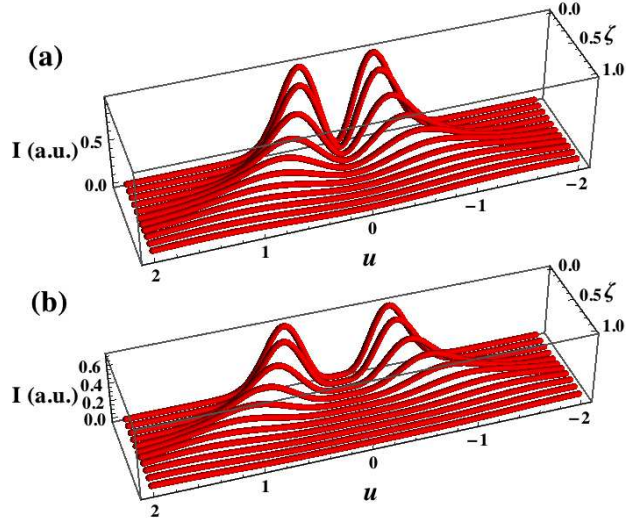}
	\caption{\label{fig:propagation} (Color online) Simulated propagation of two different RCS: (a) $m=1$, and (b) $m=2$ with $\alpha=0.6$.}
\end{center}
\end{figure}
(i) \textit {Radial coherent state}: the ground state of radial nodes is $\ket{\mbox{0,m}}$, so, the CS is
\begin{eqnarray}\label{eq:gcs_def}
	\ket{\alpha}_m=e^{\left(\alpha\op{\cal{P}}_+-\alpha^{\star}\op{\cal{P}}_-\right)}\ket{\mbox{0,m}},
\end{eqnarray}
where $\alpha=\tanh{(\xi/2)}\,e^{i\theta}$ is a complex number and is given in terms of the two real quantities $(\xi,\theta)$. The coherent parameter for su(1,1) is bounded by the unit Poincar\'e disk, i.e. $|\alpha|<1$. We may expand the CS in terms of the LG modes:
\begin{eqnarray}\label{eq:gcs_lg_exp}
	 \ket{\alpha}_m=(1-|\alpha|^2)^{\left(\frac{|m|+1}{2}\right)}\sum_{p=0}^{\infty}\sqrt{\frac{\left(p+|m|\right)!}{|m|!p!}}\,\alpha^p\ket{p,m},
\end{eqnarray}
where $|m|!$ is a normalization factor. The radial CS (RCS) depends on the OAM value as well. Therefore, for each values of the OAM, there will be a specific set of RCS. The RCS is not an orthogonal set of modes: ${}_{m'}\!\braket{\alpha'}{\alpha}_{m}=\left((1-|\alpha|^2)(1-|\alpha'|^2)/(1-\alpha\alpha'^{\ast})^2\right)^{(|m|+1)/2}\, \delta_{m,m'}$, nevertheless, the set is complete
\begin{eqnarray}\label{eq:gcs_completeness}
	\frac{|m|}{\pi}\int \frac{d^2\alpha}{(1-|\alpha|^2)^2}\,\ket{\alpha}_m\,{}_{m}\!\bra{\alpha}=\op{\mathbb{I}},
\end{eqnarray}
where $\frac{|m|}{(1-|\alpha|^2)^2}$ is the weight of measure. Therefore, the RCS is an over-complete set solution of the radial modes.\\
The RCS is a solution of the PWE and, hence, provides a novel type of the paraxial modes. It can be shown that $\sum_{p=0}^{\infty}|\braket{\mbox{p,m}}{\alpha}_m|^2=1$, so, the RCS carries a finite power. Fig~(\ref{fig:lg_basis}) shows the projection of different RCS in the basis of the first hundred p number of the LG modes.\\
The position representation of RCS at the pupil, $\mbox{RCS}_{\alpha,m}(\rho,\phi,0)=\braket{\rho,\phi,0}{\alpha}_m$, is given by
\begin{eqnarray}\label{eq:gcs_position_rep_pupil}
	 \mbox{RCS}_{\alpha,m}(\rho,\phi,0)=\sqrt{\frac{2^{|m|+1}}{\pi\,|m|!}}\left(\frac{1-\alpha^{\ast}}{1-\alpha}\right)^{\frac{|m|+1}{2}}e^{i\,m\phi-\frac{(1+\alpha)}{1-\alpha}\,\rho^2}\rho^{|m|}.\nonumber
\end{eqnarray}
Due to the presence of $e^{im\phi}$, photons in the RCS mode $\ket{\alpha}_m$ carry a well-defined OAM value of $m\hbar$ per photon. Propagation of two different RCSs is shown in Fig. (\ref{fig:propagation}). The mode intensity vanishes as  $\rho^{-2|m|}€$ at the beam center, so that that this mode can be generated starting from a Gaussian beam by simply combining a phase singularity of charge $m$ and parabolic transmission filter of order $|m|$. When $\alpha=0$ the RCS reduces to a subfamily of (type-I) HyGG modes with $p=0$ (the first class of the so-called modified LG modes)~\cite{karimi:07}. As discussed in Ref.~\cite{karimi:07}, the HyGG modes are over-complete set of modes yet non-orthogonal solutions of the PWE as well.

(ii) \textit{Radial intelligent state}: the eigenvector of lowering operator is named the IS. Here we are seeking for a special radial intelligent state (RIS), which obeys the following eigenvalues problem
\begin{eqnarray}\label{eq:intelligent_def}
	\op{\cal{P}}_-\ket{\eta}_m=\eta\ket{\eta}_m.
\end{eqnarray}
\begin{figure}[h]
\begin{center}
	\includegraphics[scale=0.33]{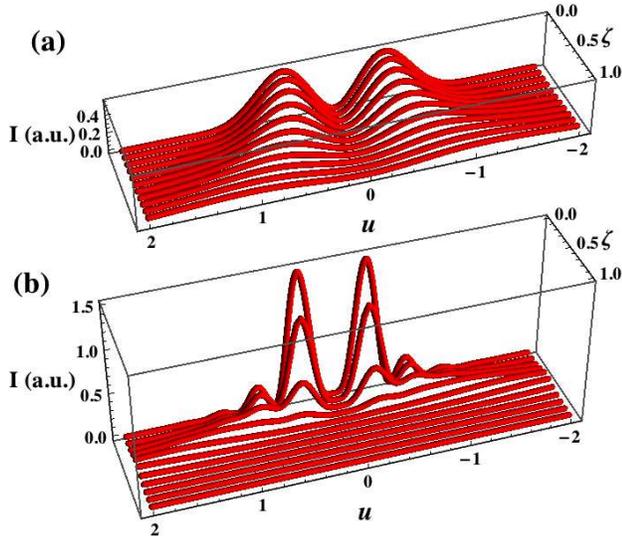}
	\caption{\label{fig:prop_bg} (Color online) Simulated propagation of two different RISs: (a) $m=1$ with $\eta=1$ and (b) $m=2$ with $\eta=10$, respectively. The later case has several rings at the pupils, which during propagation collapses to a doughnut shape.}
\end{center}
\end{figure}
Equation (\ref{eq:intelligent_def}) can be solved in the $\ket{\mbox{p,m}}$ basis yielding
\begin{eqnarray}\label{eq:intelligent_exp}
	 \ket{\eta}_m=\frac{\eta^{\frac{|m|}{2}}}{\sqrt{I_{|m|}\left(2|\eta|\right)}}\sum_{p=0}^{\infty}\frac{\eta^p}{\sqrt{p!\left(p+|m|\right)!}}\,\ket{\mbox{p,m}}
\end{eqnarray}
where $\eta$ is a free complex parameter, and $I_{n}(.)$ is the modified Bessel function of order $n$. Like the RCS modes, the RIS are a not orthogonal set of modes, since we have  ${}_{m'}\!\braket{\eta'}{\eta}_{m}=I_{|m|}\left(2\sqrt{\eta\,\eta'^{\ast}}\right)/\sqrt{\left(I_{|m|}\left(2|\eta|\right)I_{|m|}\left(2|\eta'|\right)\right)} \delta_{m,m'}$. Nevertheless, they are a complete set since we have
\begin{eqnarray}\label{eq:is_completeness}
	\frac{2}{\pi}\int {d^2\eta}\,K_m\left(2|\eta|\right) I_m\left(2|\eta|\right)\,\ket{\eta}_m\,{}_{m}\!\bra{\eta}=\op{\mathbb{I}},
\end{eqnarray}
where $K_{n}(.)$ is the $n$-order modified Bessel function of the second kind. In order to have an easy and clear picture of the RIS, we may calculate its representation in the position space, i.e. $\mbox{RIS}(\rho,\phi,0)=\braket{\rho,\phi,0}{\eta}_m$
\begin{eqnarray}\label{eq:ris_position_rep_pupil}
	\mbox{RIS}_{\eta,m}(\rho,\phi,0)=\sqrt{\frac{2}{\pi\, I_{|m|}(2|\eta|)}}\,e^{im\phi+\eta-\rho^2}\,J_{|m|}\left(2\sqrt{2\eta}\rho\right).\nonumber
\end{eqnarray}
We see that the RIS is the same as the so-called Bessel-Gauss beam. The Bessel-Gauss beam is then the intelligent state of radial modes and is the eigenstate of the lowering operator in the LG base representation. Like the RCS, the RIS is an eigenstate of the light OAM as well. Nevertheless, unlike RCS, it is not shape invariant under free-air propagation for general parameter $\eta$. Fig. (\ref{fig:prop_bg}) shows the propagation of two RIS beams with different $\eta$. We see that the intensity of one of the beams is fraught with a dramatic change during propagation.

In conclusion, we have shown that the dynamical symmetry of the radial node index of the scalar PWE is related to the su(1,1) Lie algebra. Based on this algebra, we studied two kinds of CS associated to the radial profile. Such CSs, indeed, may be used for minimizing the uncertainty relation between radial index number and its conjugate variable. 

We acknowledge the financial support of the Future and Emerging Technologies (FET) programme within the Seventh Framework Programme for Research of the European Commission, under FET-Open grant number 255914- PHORBITECH.


\begin{thebibliography}{99}


\bibitem{arnold:08}  
S.~Franke-Arnold, L.~Allen, and M.~J.~Padgett,  Laser Photonics Rev.\, {\bf 2}, 299, (2008).

\bibitem{gibson:04} 
G.~Gibson, J.~Courtial, M.~J. Padgett, M.~Vasnetsov, V.~Pasko, S.~M. Barnett,  and S.~Franke-Arnold, \newblock \opex {\bf 12}, 5448 (2004).

\bibitem{mair:01} 
A.~Mair, A.~Vaziri, G.~Welhs, and A.~Zeilinger, \newblock \nat {\bf 412}, 313 (2001).


\bibitem{nagali:09b} 
E.~Nagali, L.~Sansoni, F.~Sciarrino, F.~De Martini, L.~Marrucci, B.~Piccirillo, E.~Karimi, and E.~Santamato, \newblock Nat.\ Photon. {\bf 3}, 712 (2009).

\bibitem{karimi:10a} 
E.~Karimi, S. Sluserenko, B.~Piccirillo, L.~Marrucci, and E.~Santamato, \newblock \pra {\bf 81}, 053813 (2010).

\bibitem{padgett:04} 
M.~Padgett, J.~Courtial, and L.~Allen, \newblock Physics Today {\bf 57}, 35 (2004).

\bibitem{berry:04} 
M.~V.~Berry, \newblock J.\ Optics.\ A {\bf 6}, 259 (2004).

\bibitem{bekshaev:08}
A.~Ya.~Bekshaev, A.~I.~Karamoch, \newblock \oc  {\bf 281} 1366 (2008).

\bibitem{karimi:07} 
E.~Karimi, G.~Zito, B.~Piccirillo, L.~Marrucci, and E.~Santamato, \newblock \ol {\bf 32}, 3053 (2007).


\bibitem{karimi:09a} 
E.~Karimi, B.~Piccirillo, L.~Marrucci, and E.~Santamato, \newblock \ol {\bf 34}, 1225 (2009).

\bibitem{karimi:09b} 
E.~Karimi, B.~Piccirillo, E. Nagali, L.~Marrucci, and E.~Santamato, \newblock \apl {\bf 94}, 231124 (2009).

\bibitem{molinaterriza:07a} 
G.~Molina-Terriza, L.~Rebane, J.~P.~Torres, L.~Torner, S.~Carrasco, \newblock J.\ Europ.\ Opt.\ Soc.\ Rap.\ Public. {\bf 2}, 07014 (2007).

\bibitem{perelomov:86}
A. Perelomov, {\it Generalized coherent states and their applications}, Springer(1986).


\bibitem{schrodinger:26}
E. Schr\"odinger , Naturwissenschaften {\bf 14}, 664 (1926).

\bibitem{barut:71}
A.~O.~Barut, L.~Girardello, Commun.\ Math.\ Phys.\  {\bf 21}, 41 (1971).

\bibitem{note}
The following dimensionless cylindrical and cartesian coordinates have been used $(\rho=r/w_0,\phi,\zeta=z/z_R)$ and $(u=x/w_0,v=y/w_0,\zeta=z/z_R)$, respectively, where $w_0$ is the beam waist and $z_R=\pi{w_0}^2/\lambda$ is the beam Rayleigh range.

\bibitem{siegman:86}
A.~E.~Siegman, {\it Lasers}, University Science Books (1986).

\end{thebibliography}
\end{document}